\documentclass[twocolumn]{article}
\usepackage[T1]{fontenc}
\usepackage[utf8]{inputenc}
\usepackage{ismir,amsmath,cite,url}
\usepackage{multicol}
\usepackage{graphicx}
\usepackage{color}
\usepackage{kotex}
\usepackage{blindtext}
\usepackage{multirow}
\usepackage{placeins}
\usepackage{mwe}
\usepackage{subfig}

\title{Exploring Aligned Lyrics-Informed Singing Voice Separation}

\oneauthor
  {Chang-Bin Jeon, Hyeong-Seok Choi and Kyogu Lee}
{Department of Intelligence and Information \\ Music and Audio Research Group (MARG) \\ Center for Superintelligence \\ Seoul National University \\ { \tt \{vinyne, kekepa15, kglee\}@snu.ac.kr}}

\begin{document}

\maketitle

\begin{abstract}
In this paper, we propose a method of utilizing aligned lyrics as additional information to improve the performance of singing voice separation.
We have combined the highway network-based lyrics encoder into \textit{Open-unmix} separation network and show that the model trained with the aligned lyrics indeed results in a better performance than the model that was not informed.
The question now remains whether the increase of performance is actually due to the phonetic contents that lie in the informed aligned lyrics or not.
To this end, we investigated the source of performance increase in multifaceted ways by observing the change of performance when incorrect lyrics were given to the model.
Experiment results show that the model can use not only just vocal activity information but also the phonetic contents from the aligned lyrics.
\end{abstract}

\section{Introduction}\label{sec:introduction}

Singing voice separation is one of the most widely studied areas in the field of audio signal processing. In particular, the importance of research is greatly emphasized because it can contribute to the pre-processing step of research in various fields of Music Information Retrieval (MIR), such as automatic music transcriptions and automatic lyrics alignments. With the recent development of deep neural networks, a number of music source separation studies have been published and showed excellent performance.
These studies have a common feature of separating music source by using only information from the sound source itself, such as a 1-dimensional waveform \cite{stoller2018wave, defossez2019music} or a 2-dimensional spectrogram \cite{jansson2017singing, takahashi2018mmdenselstm, stoter19}.

One of the distinguishing characteristics that differentiate music signals from other audio signals is that usually there exists the corresponding music scores or lyrics. 
Therefore, several studies attempted to separate the sources by utilizing additional information other than the information of the sound source itself. For example, the additional information such as pitch \cite{cano2014pitch} or even the whole score \cite{woodruff2006remixing} can be used as a prior to help separate the source of interest from the mixture. 
In general, however, music scores for certain songs are not readily available, while lyrics can be easily collected on the web.

The lyrics are particularly closely related information to singing voice; thus, have promising possibility to be used as additional information for singing voice separation.
Recent studies proposed the ways of using linguistic features extracted from the end-to-end automatic speech recognition model \cite{takahashi2020improving} or voice conversion model \cite{chandna2020content} to the singing voice separation framework.
However, the way of using explicit lyrics information has not been studied enough so far, which motivates us to study the possibility of lyrics-informed singing voice separation.
We expect that singing voice separation systems can benefit from lyrics information because of the rich information contained in the phonetic features such as formant frequencies.

To utilize the lyrics, we combined the highway network-based lyrics encoder \cite{lee2019adversarially} into the current state-of-the-art music source separation network, \textit{Open-unmix} \cite{stoter19}.
In addition, we tried two conditioning methods --- 1. local conditioning, 2. concatenation --- and compare the performance.
Note that the alignment between the lyrics and songs itself is another separate line of research \cite{gupta2019acoustic, stoller2019end}. For our study, we assume that the alignment is already done and only focus on the use of the \textit{aligned} lyrics.

The information in the aligned lyrics can be seen from two perspectives: 1. The timing information of vocal activity, 2. phonetic information.
Therefore, it is important to check if the network is using the phonetic information other than the vocal activity information.
Various evaluations were conducted to examine whether the phonetic information of the aligned lyrics actually contribute to improving the performance.
We found that the proposed model trained with the aligned lyrics clearly show better performance than the baseline model trained without any additional information. 
Furthermore, the experiment results show that the performance of the proposed model even exceeds the model trained only with additional vocal activity information.
To the best of our knowledge, this is the first research to directly use the lyrics information for a singing voice separation task.

\section{Related work}\label{sec:related work}
\subsection{Informed Source Separation}
Several studies using machine learning algorithms other than deep neural networks attempted to separate singing voice with side information. 
For example, robust Principal Component Analysis (rPCA) was used
with additional vocal activity information \cite{chan2015vocal}. Also, rPCA and Non-negative Matrix Factorization (NMF) were used with
pronounced lyrics  \cite{chen2013spoken}.

Although only few studies have tried to use additional information for singing voice separation using deep neural networks, it was reported that vocal activity information can be used as an input to the network along with spectrogram to enhance the performance of the singing voice separation network \cite{schulze2019weakly}. Very recently, in the speech enhancement field, attempts have been made to utilize text information to increase the separation performance \cite{schulze2020joint}.

\begin{figure}
 \centerline{\includegraphics[width=4cm]{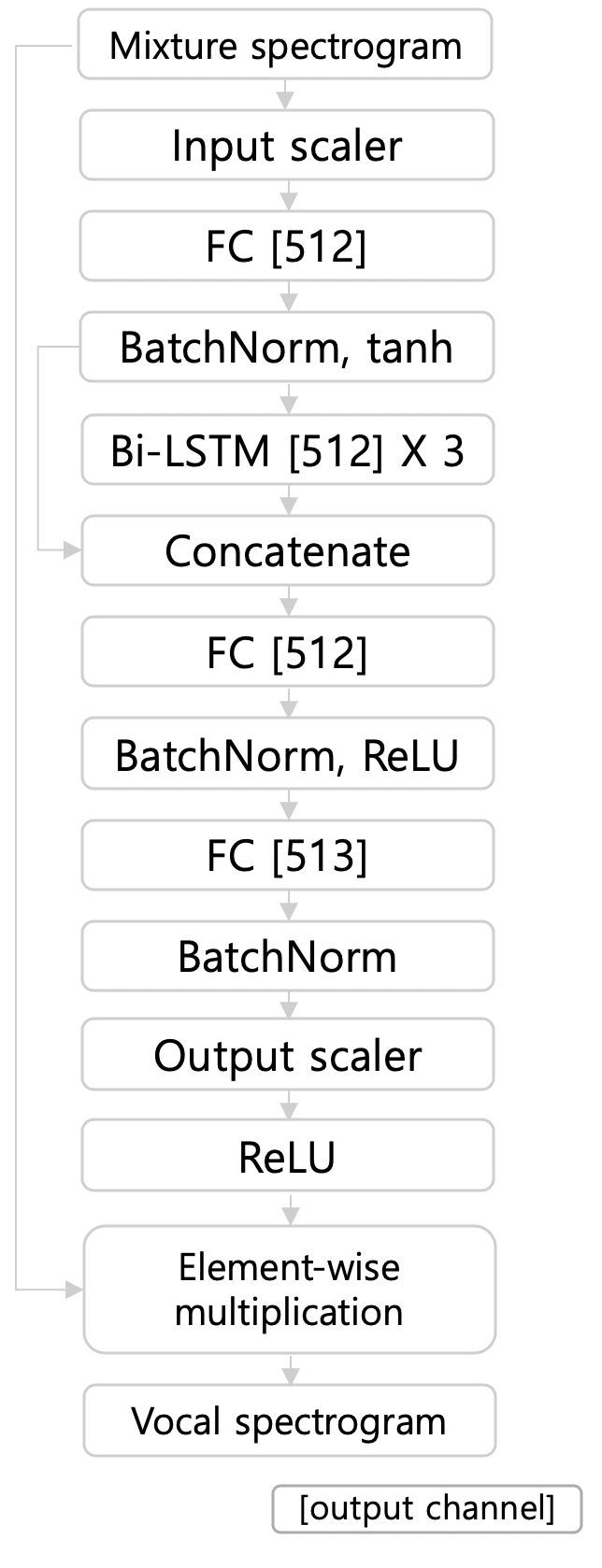}}
 \caption{The structure of the baseline \textit{Open-unmix} network.}
 \label{fig:umx_baseline}
\end{figure}

\subsection{Open-unmix}\label{subsec:open-unmix}

\textit{Open-unmix} \cite{stoter19} is the state-of-the-art music source separation network using the MUSDB18 dataset \cite{MUSDB18}. In particular, it consists of 3 bi-directional Long Short-Term Memory (LSTM) layers for source separation with 3 additional fully-connected layers. Batch normalization \cite{ioffe2015batch} was used after every fully-connected layer and skip connection \cite{he2016deep} was used between the inputs and outputs of 3 consecutive bi-directional LSTM layers. Trainable input and output scalers through frequency-axis are also the special features of \textit{Open-unmix}, differentiating from other studies that use decibel scales.

We used \textit{Open-unmix} network as our baseline model because we wanted to check whether the aligned lyrics information could improve the performance of the current state-of-the-art model. The channel inputs and outputs are mono in our study although the original study used stereo. 
This is because our singing dataset is made up of a clean singing voice without any reverberation, chorus, or doubling, as opposed to the singing tracks in MUSDB18's singing voice dataset, which are already processed for the stereo.
Details and the full structure of the baseline \textit{Open-unmix} are illustrated in \figref{fig:umx_baseline}.

\begin{figure}
 \centerline{\includegraphics[width=3.75cm]{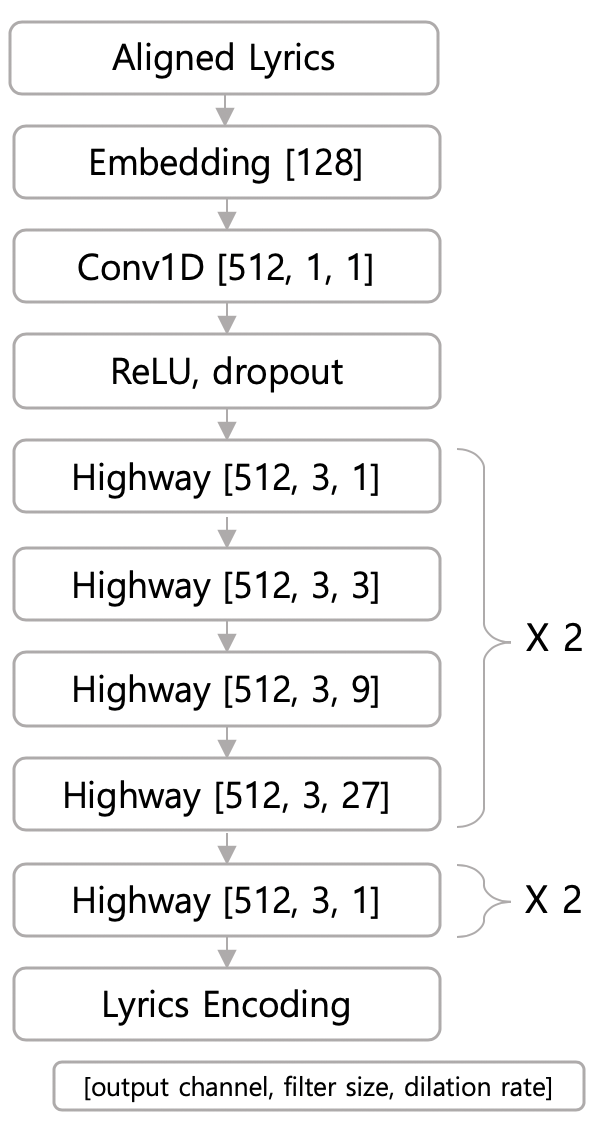}}
 \caption{The structure of the lyrics encoder.}
 \label{fig:lyrics_encoder}
\end{figure}

\subsection{Singing Voice Synthesis}\label{subsec:ksvs network}
Singing voice synthesis and lyrics-informed singing voice separation tasks share a similar framework in that the input and output are the same as lyrics and singing voice spectrograms, respectively.
Recently, \cite{lee2019adversarially} proposed the singing voice synthesis network that succeeds in creating high-quality singing based on 60 Korean songs sung by a single singer. 
It is based on the Text-to-Speech model \cite{tachibana2018efficiently}, which consists of 1-dimensional convolutional neural networks and highway networks \cite{srivastava2015highway}.
Therefore, we borrowed the idea of using the highway network-based lyrics encoder and integrated it into the source separation network.

\section{Proposed Method}\label{sec:proposed method}

\subsection{Lyrics Encoder}\label{subsec:body}
The detailed structure of the lyrics encoder is shown in \figref{fig:lyrics_encoder} and the structure of the highway networks used in the lyrics encoder is defined as follows,
\begin{equation} \label{highway}
	y = ReLU (x \ast W_{H}) \cdot \sigma (x \ast W_{T}) + x \cdot (1 - \sigma (x \ast W_{T})),
    \end{equation}
, where $x$ and $y$ are input and output of the network and $\cdot$ refers to the element-wise multiplication. $x \ast W_{H}$ and $x \ast W_{T}$ are the 1-dimensional convolution layers which have the same input and output channel sizes. Biases are omitted in \eqnref{highway}. Zero-padding was applied to keep the input and output length the same in every convolution layer. Dropouts  \cite{srivastava2014dropout} with 0.05 dropout rate, were applied after the activation functions. Dilated convolution \cite{yu2015multi} were used to expand the receptive field to 165 frames. It is about seven times larger than the model without the dilation.
In our experimental settings, 165 frames are equal to about 1.915 seconds.
    
\begin{figure}
\begin{minipage}{.49\linewidth}
\centering
\subfloat[]{\label{main:a}\includegraphics[width=\columnwidth]{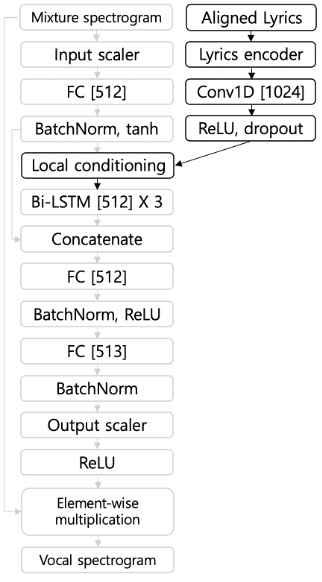}}
\end{minipage}
\begin{minipage}{.49\linewidth}
\centering
\subfloat[]{\label{main:b}\includegraphics[width=\columnwidth]{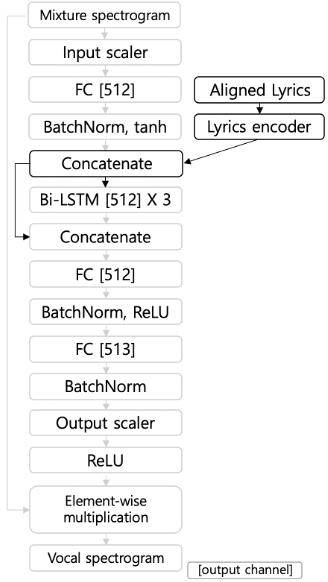}}
\end{minipage}\par\medskip
\centering
 \caption{The structure of the \textit{Open-unmix} networks combined with the lyrics encoder using (a) local conditioning and (b) concatenation method.}
 \label{fig:umx_proposed}
\end{figure}

\subsection{Local Conditioning of Lyrics Encoding}

The local conditioning method \cite{lee2019adversarially} was used to insert the encoded lyrics information into the singing voice separation network. 
The local conditioning is defined as follows, 
\begin{equation} \label{localconditioning}
	y = ReLU(x \ast W_{f} + L_{1}) \cdot \sigma(x \ast W_{g} + L_{2}),
    \end{equation}
where, $x \ast W_{f}$ and $x \ast W_{g}$ are the 1-dimensional convolution layers which have same input and output channel sizes.
$L_{1}$ and $L_{2}$ are equally separated features through the channel axis from the output of the consecutive lyrics encoder and the 1-dimensional convolutional layer. The 1-dimensional convolution layer with filter size 1 was added on the output of the lyrics encoder so that the channel size of each $L_{1}$ and $L_{2}$ could be 512. $\sigma$ refers to the sigmoid activation function. Details of the full structure are in \figref{main:a}.

\subsection{Concatenation of Lyrics Encoding}
A concatenation method, which is a simple but powerful conditioning method, was also used in our study to pass the encoded lyrics information into the singing voice separation network. By concatenating the output of the first fully-connected layer and the lyrics encoder output, the channel size of LSTM layers input becomes 1024. Also, the channel size of the second fully-connected layer input becomes 1536 by the skip-connection of LSTM input and output. Details of the full structure is in \figref{main:b}.

\section{Experiments}
\label{sec:experiments}

\subsection{Dataset}
Here we used a total of 201 Korean pop songs sung by 13 amateur singers as target clean singing sources. 
This dataset has a total length of 11 hours and 44 minutes. Of these, we used 162 songs (9h 3m) for training, 19 songs (1h 7m) for validation, and 20 songs (1h 7m) for test dataset.
The detailed composition of the dataset is described in \tabref{tab:dataset}.

We aligned Korean syllable following \cite{lee2019adversarially}; one Korean syllable is made up of onset (consonant), nucleus (vowel) and coda (consonant), we aligned onset and coda for 4 frames, and nucleus for other frames.
The example of the alignment applied to the spectrogram is shown in \figref{fig:baram}.

\begin{table}

\begin{center}
\resizebox{6.5cm}{!}{
    \begin{tabular}{|c||c|c|c|c||c|}
    \hline
    Singer & Gender & Train  & Validation & Test & Total   \\\hline\hline
    1   & Female  & 79  & 5  & 8  & 92  \\ \hline
    2   & Male  & 8   & 2  & 0  & 10  \\ \hline
    3   & Female  & 8   & 2  & 0  & 10  \\ \hline
    4   & Female  & 9   & 0  & 1  & 10  \\ \hline
    5   & Female  & 8   & 0  & 1  & 9   \\ \hline
    6   & Male  & 10  & 0  & 0  & 10  \\ \hline
    7   & Female  & 8   & 0  & 2  & 10  \\ \hline
    8   & Female  & 9   & 1  & 0  & 10  \\ \hline
    9   & Male  & 9   & 0  & 1  & 10  \\ \hline
    10  & Male  & 7   & 1  & 1  & 9   \\ \hline
    11  & Female  & 7   & 3  & 0  & 10  \\ \hline
    12  & Female  & 0   & 5  & 5  & 10  \\ \hline
    13  & Male  & 0   & 0  & 1  & 1   \\ \hline\hline
       &    & 162 & 19 & 20 & 201 \\ \hline
    \end{tabular}
    }
    \end{center}
    \caption{The composition of our singing dataset.}
    \label{tab:dataset}
    \end{table}
    
A total of 19,113 instrumental songs were used as accompaniment for training networks because we did not have real accompaniment tracks corresponding to the singing voice dataset. Since various studies using MUSDB18\cite{MUSDB18} or DSD100 \cite{liutkus20172016} dataset also used random mixing techniques, i.e. creating random accompaniment for each iteration that was not related to the original singing, we decided that using arbitrary accompaniments would not be a problem for training. In addition, if we used the same specific accompaniments for the test singing voice dataset, we assumed that it is reasonable for identifying how the information containing the phonetic features of the aligned lyrics has changed the performance of the network, the fact we wanted to identify. Therefore, for the validation and test dataset, we randomly chose each 19 and 20 instrumental songs which have a longer length than the singing data, and shorten the length to the same with the singing.

\begin{figure}
 \centerline{\includegraphics[width=\columnwidth]{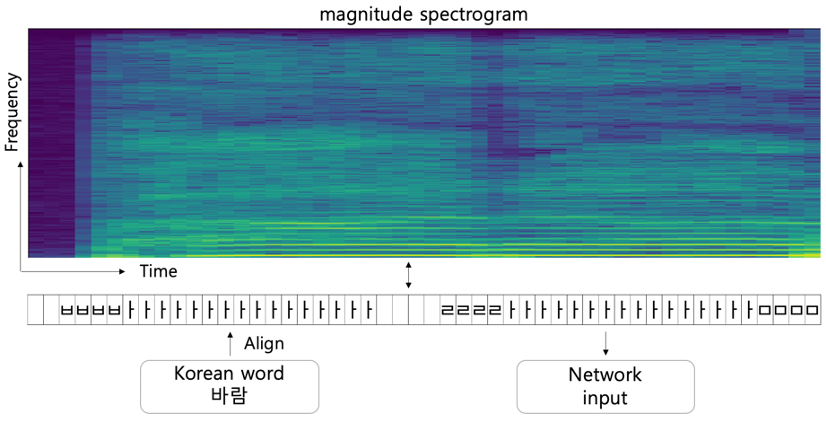}}
 \caption{The example of inserting the aligned lyrics to the networks.}
 \label{fig:baram}
\end{figure}

\subsection{Training}
In our singing dataset, the number of songs recorded by the first singer is outnumbered compared to the others, accounting for about 46 percent of the total. In order to prevent bias to a particular singer when training the networks, a singer was first selected with the same probability when constructing a batch to be used for each iteration, and the vocal source to be used for training was sampled only for the songs recorded by the selected singer.

A mono sound source with a sample rate of 22050 Hz was used in the experiment. FFT point size and window size were set to 1024 samples (0.0464 seconds) to convert them into a spectrogram, and Short-time Fourier transform (STFT) hop size to 256 samples (0.0116 seconds). Adam optimization method \cite{kingma2014adam} was used for training with a learning rate of 0.001, $\beta_{1}$ for 0.9, $\beta_{2}$ for 0.999. Mean squared error (MSE) loss function between the ground-truth and the outputs of the models were used in our study. We trained the models for 500 epochs with calculating validation loss for every epoch. The learning rate was reduced to 30 percent if there was no decrease in validation loss during 25 epochs. Early stopping was applied after 50 epochs without a decrease in validation loss.

\subsection{Evaluation Methods}
Here we briefly summarize the various experiments that will be shown in the following Section \ref{sec:results}.
The experiments will be conducted in three following ways.

First, in Section \ref{subsec:performance}, we compare and analyze how much performance improvement there are between the baseline model trained without the lyrics and the model trained with the aligned lyrics.

Second, in Section \ref{subsec:va_usage}, we check if the network exploits the vocal activity information included in the aligned lyrics. The lyrics include both the vocal activity information and phonetic information, and thus expected to use the vocal activity information correctly.

Third, in Section \ref{subsec:incorrect}, we check if the given input is correctly being used by the network trained with aligned lyrics.
This experiment was done by giving incorrect inputs in the evaluation stage.
It is expected that the incorrect inputs will significantly reduce performance.

For network performance evaluations, Signal-to-Distortion Ratio (SDR), Signal-to-Interference Ratio (SIR), Signal-to-Artifact Ratio (SAR) scores \cite{vincent2006performance} were computed by museval python library \cite{SiSEC18}.

\section{Results}
\label{sec:results}

\begin{table}
\small
\begin{center}
\begin{tabular}{|c|c|}
\hline
Model name & Inputs to the lyrics encoder \\ \hline\hline
\textit{model 1}    & None                         \\ \hline
\textit{model 2}    & Meaningless inputs (all 0)   \\ \hline
\textit{model 3}    & Vocal activity information   \\ \hline
\textit{model 4}    & Aligned lyrics               \\ \hline
\end{tabular}
\end{center}
\caption{The description of each models in our experiments.}
\label{tab:model_description}
\end{table}

The configuration of the models we trained in our experiments is in \tabref{tab:model_description}. 
We trained four models each using local conditioning and concatenation method. \textit{model 1} is the baseline model that trained without the lyrics encoder. \textit{model 2} is the model that trained with meaningless 0 value inputs to the lyrics encoder. This model is only for checking the performance change of the networks caused by the extended network capacity. \textit{model 3} is the model that trained with only vocal activity information to the lyrics encoder. We simply used 0 value as unvoiced sections and 1 as voiced sections so that the 128-dimensional embedding can train useful meaning from it. \textit{model 4} is that trained with aligned lyrics information. For both \textit{model 3} and \textit{model 4}, note that 0 value has clear meaning, unvoiced sections, unlike 0 value in \textit{model 2} is meaningless. Since the baseline model is the same for each local conditioning and concatenation method, we trained a total of 7 models for the experiments. Except \textit{model 1}, we will use the abbreviation of local conditioning and concatenation methods, each \textit{LC} and \textit{CC}, in front of the model names for convenience. For example, the model trained with aligned lyrics and the local conditioning method is \textit{LC-model 4}.

\begin{table*}
\centering
\begin{tabular}{|c||c|c|c||c|c|c|}
\hline
\multirow{2}{*}{Models}      & \multicolumn{3}{c||}{Median}                         & \multicolumn{3}{c|}{Mean}                         \\ \cline{2-7} 
                             & SDR             & SIR             & SAR             & SDR            & SIR             & SAR            \\ \hline\hline
\textit{model 1}& 9.956 & 18.674 & 9.847 & 8.595 & 16.062 & 9.145\\ \hline\hline
\textit{LC-model 2} & 10.140          & 18.465          & 9.766           & 8.589          & 16.001          & 9.093          \\ \hline
\textit{LC-model 3} & 10.090          & 18.713          & 9.763           & 9.250          & 16.298          & 9.153          \\ \hline
\textit{LC-model 4} & \textbf{10.767} & 19.505          & 10.223          & 9.723          & 17.116          & 9.699          \\ \hline\hline
\textit{CC-model 2} & 10.110          & 18.434          & 9.909           & 8.691          & 16.164          & 9.207          \\ \hline
\textit{CC-model 3} & 10.444          & 19.328          & 10.169          & 9.718          & 17.031          & 9.609          \\ \hline
\textit{CC-model 4} & 10.757          & \textbf{19.623} & \textbf{10.371} & \textbf{9.752} & \textbf{17.250} & \textbf{9.803} \\ \hline
\end{tabular}
\caption{Evaluation scores of our singing voice separation models. All scores are in [dB] scale.}
\label{tab:eval_score}
\end{table*}

\subsection{Performance Evaluation}
\label{subsec:performance}

\begin{figure}
 \centerline{\includegraphics[width=.95\columnwidth]{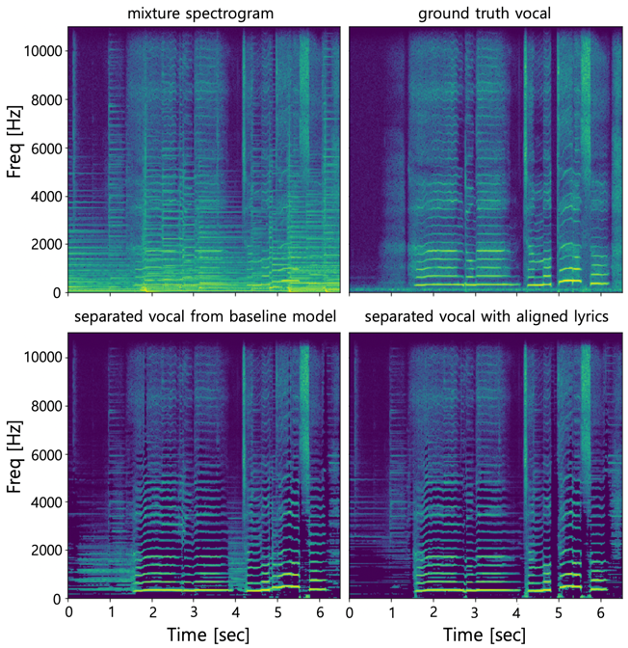}}
 \caption{The examples of the mixture, ground truth vocal, separated vocal spectrograms of baseline \textit{model 1} and \textit{CC-model 4}.}
 \label{fig:inferred_spec}
\end{figure}

The quantitative performance evaluation scores of the models are shown in \tabref{tab:eval_score}. Median scores were taken from the median values of 20 songs, which were calculated for every frame (a median of frames, a median of tracks). Mean scores represent the scores taken by a mean of frames, a mean of tracks. Each frame was set to 1 second.

It was confirmed that the separation performance of both \textit{LC-model 4} and \textit{CC-model 4} improved from the \textit{model 1}. This implies that aligned lyrics information can be used as helpful features for singing voice separation networks. Comparing to \textit{LC-model 3} and \textit{CC-model 3}, we could verify that there were clear performance gains not only from the lyrics alignment information but also the phonetic features of the lyrics itself. It was also confirmed that the performance gains do not come from just network capacity growth, given that there are no significant differences in the performance of \textit{model 1} and \textit{model 2}. The spectrograms of the separated sample are in \figref{fig:inferred_spec}.

Despite the expectation that the vocal activity information is powerful information to the networks, performance gains observed in \textit{LC-model 3} were very slight. It was much smaller than the improvements achieved from \textit{CC-model 3}. By these, we analyzed that the concatenation method is slightly better for making the networks to reflect the vocal activity information. Nevertheless, we considered that both conditioning methods were effective when giving the networks aligned lyrics information.

\subsection{Analysis of Vocal Activity Information Usage}
\label{subsec:va_usage}
In this section, we quantitatively assessed how well the networks leverage vocal activity information of aligned lyrics. The purpose of this is to see if the models have not been trained by focusing only on either one of the vocal activity information or the phonetic information of aligned lyrics, which are both critical for the separation performance.

To this end, the separated spectrogram values were divided by the largest values of each source for normalization, so that the minimum and maximum values become 0 and 1. Then, the energy of each time axis was summed to create a vector that contains vocal activity information. It was decided whether vocal activity exists or not based on whether the values were larger or smaller than 0.1 as was done in \cite{schulze2019weakly}. Precision, recall, and F1 scores were calculated with the created vocal activity vectors by taking the place where the lyric exists as the ground-truth voiced sections. Scores are contained in \tabref{tab:eval2}.

\begin{figure}
 \centerline{\includegraphics[width=\columnwidth]{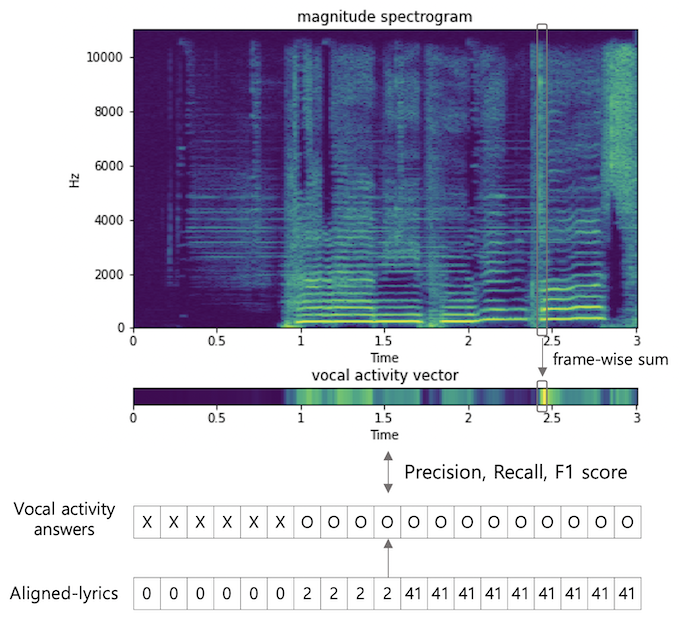}}
 \caption{The example of making the vocal activity vector from the separated vocal spectrogram.}
 \label{fig:baram}
\end{figure}

From the results of \tabref{tab:eval2}, we have confirmed that \textit{model 4} can separate the vocal source by reflecting the vocal's timing information more accurately than \textit{model 1} for both lyrics conditioning methods. Also, it was confirmed that \textit{model 3} achieved higher scores for all measures than \textit{model 4}. This is a reasonable result because \textit{model 3} were trained with vocal activity information only, while \textit{model4} needed to learn how to leverage both vocal activity information and phonetic information appropriately while training. Nevertheless, F1 score differences were negligible, which means \textit{model 4} was also capable of reflecting timing information as well as \textit{model 3}.

\subsection{Analysis with Using Incorrect Lyrics}
\label{subsec:incorrect}
To check if \textit{model 4} effectively uses the information in lyrics, we observed the performance change when incorrect lyrics were given as input during the evaluation stage. The results are shown in \tabref{tab:eval3_1}.

\begin{table}
\begin{center}
\begin{tabular}{|c||c|c|c|}
\hline
Models              & Precision      & Recall         & F1 score       \\ \hline\hline
\textit{model 1}    & 0.807          & 0.853          & 0.828          \\ \hline\hline
\textit{LC-model 2} & 0.810 & 0.852          & 0.830          \\ \hline
\textit{LC-model 3} & 0.887 & \textbf{0.857} & 0.872          \\ \hline
\textit{LC-model 4} & 0.876 & 0.854          & 0.865          \\ \hline\hline
\textit{CC-model 2} & 0.814 & 0.853          & 0.833          \\ \hline
\textit{CC-model 3} & \textbf{0.896} & 0.855 & \textbf{0.875} \\ \hline
\textit{CC-model 4} & 0.879 & 0.855 & 0.867 \\ \hline
\end{tabular}
\end{center}
\caption{The precision, recall, and F1 scores to evaluate how well the networks used the vocal activity information from aligned lyrics.}
\label{tab:eval2}
\end{table}

\begin{table}[]
\small
\begin{center}
\resizebox{8cm}{!}{\begin{tabular}{|c|c||c|c|c|}
\hline
Models                      & Inputs         & SDR    & SIR    & SAR    \\ \hline\hline
\multirow{4}{*}{\textit{LC-model 4}} & \texttt{Zero}           & 0.001  & 8.040  & -4.286 \\ \cline{2-5} 
                            & \texttt{Random}         & 5.317  & 15.802 & 4.845  \\ \cline{2-5} 
                            & \texttt{VA+Random}    & 7.899  & 19.270 & 6.403  \\ \cline{2-5} 
                            & \texttt{AlignedLyrics} & \textbf{10.767} & 19.505 & 10.223 \\ \hline\hline
\multirow{4}{*}{\textit{CC-model 4}} & \texttt{Zero}           & 0.002  & 7.246  & -3.671 \\ \cline{2-5} 
                            & \texttt{Random}         & 0.946  & 14.837 & 0.341  \\ \cline{2-5} 
                            & \texttt{VA+Random}    & 7.164  & 19.545 & 6.290  \\ \cline{2-5} 
                            & \texttt{AlignedLyrics} & 10.757 & \textbf{19.623} & \textbf{10.371} \\ \hline
\end{tabular}
}
\end{center}
\caption{Performance comparisons when different inputs are given in the evaluation stage. \texttt{Zero} : 0 value inputs. \texttt{Random} : Random value inputs. \texttt{VA+Random} : Replace voiced sections with random value. \texttt{AlignedLyrics} : Aligned lyrics (The proposed method). All scores are in [dB] scale.}
\label{tab:eval3_1}
\end{table}

\begin{figure}[!t]
 \centerline{\includegraphics[width=.97\columnwidth]{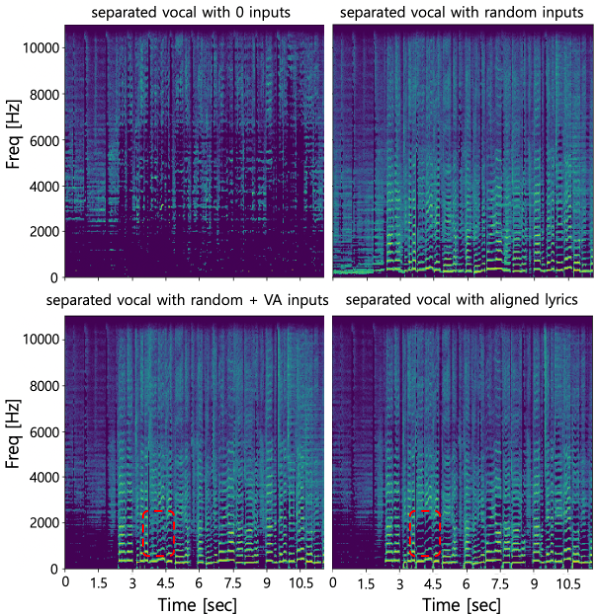}}
 \caption{The examples of the separated vocal spectrograms with incorrect inputs and correct aligned lyrics inputs were given to \textit{LC-model 4}. The dashed line shows the enhanced parts when the aligned lyrics are used. Note that the region is closely related to the formant frequencies.}
 \label{fig:inferred_zero}
\end{figure}

If the networks had learned to effectively use the information in lyrics it is expected to output silence when the lyrics meaning unvoiced sections are given in the evaluation step.
To validate this assumption, we inserted 0 values (\texttt{Zero}) to the lyrics encoder in to \textit{model 4} in the evaluation step.
As expected, almost every sound has been erased from the mixture with only a little noise left as seen in \figref{fig:inferred_zero} and critical performance degradation, over 10 dB in SDR score, has occurred.

Furthermore, performance degradation was observed when all the lyrics were replaced with random values (\texttt{Random}).
This also shows that the network is significantly dependent on the encoded lyrics information and the proposed conditioning method is effectively applied.

Next, we experimented to see if the network is able to use the phonetic information included in the lyrics.
We show this by removing all the phonetic information from the aligned lyrics. In other words, the changed lyrics still contain the vocal activity information.
More specifically, it was done by replacing all the voiced sections with random values and leaving the unvoiced sections intact (\texttt{VA+Random}).
Interestingly, the performance was still far below than the model trained tested on aligned lyrics, which indicates that the network can reflect the phonetic information into the separation process.

It is noteworthy that the SIR scores of \texttt{VA+Random} are not much different from the aligned lyrics input (\texttt{AlignedLyrics}).
Since SIR scores are heavily related to the remained accompaniment sources of the separated singing voice, we expected that the networks were still capable of removing the accompaniments only using the vocal activity information. On the other hand, the impact on SDR and SAR scores were significant. This implies that while the network was able to erase the accompaniment well by using vocal activity information only in unvoiced sections, it was able to remove the accompaniment better by using phonetic information in voiced sections.

In the experiments of using (\texttt{Random}) and (\texttt{VA+Random}) inputs, median values of 5 different experimental tries with different random seeds were taken.

\section{Conclusion}
In this study, we proposed an integrated framework of combining the lyrics encoder into the state-of-the-art \textit{Open-unmix} separation network.
Local conditioning and concatenation methods were shown to be able to effectively condition the aligned lyrics into the singing voice separation networks. 
Through various experiments, it was confirmed that the phonetic information of aligned lyrics can contribute to the performance improvements as well as the vocal activity information. 
We plan to use the unaligned lyrics for the singing voice separation for the future works.

\section{acknowledgements}
This work was supported partly by Kakao and Kakao Brain corporations and partly by Next-Generation Information Computing Development Program through the National Research Foundation of Korea (NRF) funded by the Ministry of Science and ICT (NRF-2017M3C4A7078548).
\bibliography{ISMIRtemplate}

\end{document}